# Tunable All Electric Spin Polarizer

J. Charles[1], N. Bhandari[1], J. Wan[1], M. Cahay[1,2], and R. S. Newrock[2]

[1] School of Electronics and Computing Systems
University of Cincinnati, Cincinnati, Ohio 45221, USA

[2] Physics Department, University of Cincinnati, Cincinnati, Ohio 45221, USA

## Abstract

We propose a new device to create a tunable all-electric spin polarizer: a quantum point contact (QPC) with four gates -- two in-plane side gates in series. The first pair of gates, near the source, is asymmetrically biased to create spin polarization in the QPC channel. The second set of gates, near the drain, is symmetrically biased and that bias is varied to maximize the amount of spin polarization in the channel. The range of common mode bias on the first set of gates over which maximum spin polarization can be achieved is much broader for the four gate structure compared with the case of a single pair of in-plane side gates.

Semiconductor spintronics is one of the most promising paradigms for the development of novel devices for use in the post-CMOS era [1, 2]. The major challenge of spintronics is to avoid the use of ferromagnetic contacts or external magnetic fields and to control the creation, manipulation, and detection of spin polarized currents by purely electrical means. Some major steps towards that goal have been realized recently [3-11]. Spin-orbit coupling (SOC), which couples the electron's motion to its spin, has been envisioned as a possible tool for all-electric control and generation of spin-polarized currents. It has been shown that SOC can be used to modulate spin polarized currents by taking advantage of symmetry-breaking factors such as interfaces, electric fields, strain, and crystalline directions [5]. Recently, we showed that *lateral spin-orbit coupling* (LSOC) in InAs/InAlAs and GaAs/AlGaAs quantum point contacts (QPCs) with in-plane side gates, can be used to create a strongly spin-polarized current by purely electrical means *in the absence of any applied magnetic field* [12-15]. Non Equilibrium Green's Function (NEGF) calculations of the conductance of QPCs show that the onset of spin polarization in devices with in-plane side gates required three ingredients: (1) a LSOC induced by a lateral confining potential in the QPC; (2) an asymmetric lateral confinement; and (3) a strong electron-electron interaction [16,17]. The NEGF approach was used to study in detail the ballistic conductance of asymmetrically biased side-gated QPCs in the presence of LSOC and strong electron-electron interactions for a wide range of QPC dimensions and gate bias voltages [17]. Various conductance anomalies were predicted below the first quantized conductance plateau ($G_0=2e^2/h$); these occur because of spontaneous spin polarization in the narrowest portion of the QPC. The number of observed conductance anomalies increases with increasing aspect ratio (length/width) of the QPC constriction. These anomalies are fingerprints of spin textures in the narrow portion of the QPC [17]. The NEGF approach was also used to show the importance of impurity and dangling bond scattering on the location of the conductance anomalies [17,18]. In view of these results, it seems appropriate to find a way to control and finely tune the location of a conductance anomaly, with the goal of achieving the largest possible spin polarization for a specific QPC.

We propose the device shown in Fig. 1 as a tunable all electric spin polarizer. It consists of a QPC with two sets of in-plan side gates. In this structure, a bias asymmetry $V_{sg1} - V_{sg2}$ is applied between the two gates nearest the source to create spin polarization in the channel. An additional common mode bias $V_{sweep}$ is applied to both gates. The second set of gates, nearest the drain and separated by a gap 'd' from the first set, are biased at the same potential $V_{sg3} = V_{sg4}$. The latter is varied to control the amount of depletion in region II of the QPC. We show below that this has a profound effect on the spin polarization of the QPC. In fact, we found that the range of $V_{sweep}$ over which the spin conductance polarization, $\alpha = \frac{G_{up} - G_{down}}{G_{up} + G_{down}}$, is appreciable is much larger for this four gate structure than a similar two gate structure, i.e., a QPC with channel length = $2L_2$ + d and only a side gates at potentials $V_{sg1} + V_{sweep}$ and $V_{sg2} + V_{sweep}$.

We performed NEGF simulations of a four gate InAs QPC with the following parameters: $w_1$ = 48 nm, $l_1$ = 80 nm, $w_2$ = 16 nm, $l_2$ = 22 nm, and d = 4 nm. The details of the NEGF approach are given in Ref. [16]. The effective mass in the InAs channel was $m^* = 0.023 m_0$, where $m_0$ is the free electron mass. All calculations were performed at a temperature 4.2 K. Following Lassl et al. [19], the strength of the parameter γ in the interaction self-energy characterizing the strength of the electron-electron interaction was set equal to $3.7\hbar^2/2m^*$. The strength of the parameter β in the LSOC was set equal 200 Å$^2$. The potential at the source was set equal to 0V and the one at the drain, $V_d$ to 0.3 mV in all simulations. An asymmetry in the potential of the leftmost side gates in Fig.1 was introduced by taking $V_{sg1}$ = -0.2 V + $V_{sweep}$ and

$V_{sg2} = 0.2$ V $+ V_{sweep}$. The potential on the other two gates, $V_{sg3} = V_{sg4}$ biased with a constant voltage. The conductance of the constriction was studied as a function of the sweeping (common mode) potential $V_{sweep}$. The Fermi energy was equal to 106.3 meV in the source contact and 106 meV in the drain contact, ensuring single-mode transport through the QPC. At the interface between the narrow portion of the QPC and vacuum, the conduction band discontinuities at the bottom and the top interface were modeled, respectively, as

$$\Delta E_c(y) = \frac{\Delta E_c}{2}\left[1+\cos\frac{\pi}{d}\left(y-\frac{w_1-w_2}{2}\right)\right], \quad (1)$$

and

$$\Delta E_c(y) = \frac{\Delta E_c}{2}\left[1+\cos\frac{\pi}{d}\left(\frac{w_1+w_2}{2}-y\right)\right], \quad (2)$$

to achieve a smooth conductance band change, where $d$ was selected equal to 1.6nm to represent a gradual variation of the conduction band profile from the inside of the quantum wire to the vacuum region. A similar grading was also used along the walls going from the wider part of the channel to the central constriction of the QPC. This gradual change in $\Delta E_c(y)$ is responsible for the LSOC that triggers the spin polarization of the QPC in the presence of an asymmetry in $V_{sg1}$ and $V_{sg2}$. $\Delta E_c$ in Eqns. (1) and (2) was set equal to 4.5 eV.

Figure 2 shows the conductance versus $V_{sweep}$ of the four gate QPC, where $G_T$ is the total conductance and $G_{up}$ and $G_{down}$ the contributions from the up spin and down spin electrons, respectively. The biases on the different gates are $V_{sg1} = -0.2$ V $+ V_{sweep}$, $V_{sg2} = +0.2 + V_{sweep}$, and

$V_{sg3} = V_{sg2} = 0.0$ V. Figure 2 shows a conductance anomaly around 0.5 $G_0$ (the oscillations are due to multiple reflections at the ends of the QPC [17]). It is a signature of spin polarization in the channel as $G_{up}$ is much larger than $G_{down}$ over a range of $V_{sweep}$ from -2 V to 0V. For a QPC with only one set of side gates and length = 2 $l_2$ + d = 48 nm, with $V_{sg1}$ = -0.2 V + $V_{sweep}$ and $V_{sg2}$ = +0.2 + $V_{sweep}$, the range of $V_{sweep}$ over which $G_{down} \neq G_{up}$ was found to be from -1V to 0V, half the range for the four gate QPC structure. For a QPC with length $l_2$ = 22 nm, with $V_{sg1}$ = -0.2 V + $V_{sweep}$ and $V_{sg2}$ = +0.2 + $V_{sweep}$, we found $G_{down} = G_{up}$ for all $V_{sweep}$ and there is no spin polarization in the channel.

Figure 3 is a plot of the total conductance versus $V_{sweep}$ of the four gate QPC described above with $V_{sg1}$ = -0.2 V + $V_{sweep}$, $V_{sg2}$ = +0.2 + $V_{sweep}$. The three different curves correspond to the cases where the bias on the two gates close to the drain is $V_{sg3} = V_{sg4}$ = 0.0 V, -0.1 V, and -0.2V, respectively. These curves show that the conductance anomaly can be finely tuned by varying the negative potential on gates 3 and 4. The range of $V_{sweep}$ anomaly over which the conductance anomaly appears also strongly depends on the negative bias on gates 3 and 4. This is best illustrated by plotting the spin conductance polarization α corresponding to the three curves shown in Fig.3. Figure 4 shows that the range of $V_{sweep}$ over which α is different from zero increases from 0.2 to 0.5 V when to $V_{sg3} = V_{sg4}$ is changed from 0.0V to -0.2V. In Fig.4, the maximum value of α was found nearly the same (~0.97) for $V_{sg3} = V_{sg4}$ equal to 0.0 V, -0.1 V, and -0.2V. At the value of $V_{sweep}$ corresponding to $\alpha_{max}$, the total charge density over the first half of the dual QPC (region I in Fig.1) was found equal to 0.14, 0.17, and .23x10$^{12}$ cm$^{-2}$. A qualitative explanation for these results is that a negative bias on gates 3 and 4 has a twofold effect. First, it creates a potential barrier to electrons flowing through the first portion of the four gate QPC. This enhances space-charge effects in the d region, leading to an increase in electron-electron interactions which is one of the ingredients needed for enhanced spin polarization in the channel [12,16,17]. Second, because of the proximity of gates 1-2 to gates 3-4 the electrostatic potential of gates 3 and 4 affect the conduction band profile in region I including the potential walls between the QPC channel and etched regions. Since the amount of LSOC is directly related to the slopes of the QPC walls, the proximity of gates 3-4 allows a fine tuning of the amount of spin polarization in the overall structure.

**Acknowledgment**

This work is supported by NSF Awards ECCS 0725404 and 1028483. James Charles acknowledges support under NSF-REU award 007081.

**Figure Captions**

**Fig. 1**: Four gate QPC configuration. In the simulations, the potential on the two side gates closest to the source is set equal to $V_{sg1}$ = -0.2 V + $V_{sweep}$ and $V_{sg2}$ = +0.2 + $V_{sweep}$; we used Vs = 0.0 and Vd =0.3 mV. The potentials on the two side gates closest to the drain are kept at the same fixed value. The current flows in the x-direction.

**Fig. 2:** Conductance versus $V_{sweep}$ of a four gate QPC with the following parameters: $w_1$ = 48 nm, $l_1$ = 80 nm, $w_2$ = 16 nm, $l_2$ = 22 nm, d = 4 nm. $G_T$ is the total conductance and $G_{up}$ and $G_{down}$ the contributions from the up spin and down spin electrons, respectively. The biases on the different gates are set as follows: $V_{sg1}$ = -0.2 V + $V_{sweep}$, $V_{sg2}$ = +0.2 + $V_{sweep}$, and $V_{sg3}$ = $V_{sg2}$ = 0.0 V. Also shown for comparison is $G_T$ for a QPC with only two gates with: $V_{sg1}$ = -0.2 V + $V_{sweep}$, $V_{sg2}$ = +0.2 + $V_{sweep}$ for $l_2$ = 22 and 48 nm.

**Fig. 3:** Total conductance versus $V_{sweep}$ of a four gate QPC with the following parameters: $w_1$ = 48 nm, $l_1$ = 80 nm, $w_2$ = 16 nm, $l_2$ = 22 nm, d = 4 nm. The biases on the gates closest to the source are set as follows: $V_{sg1}$ = -0.2 V + $V_{sweep}$, $V_{sg2}$ = +0.2 + $V_{sweep}$. The biases on the two gates closest to the drain are equal: $V_{sg3}$ = $V_{sg4}$ = 0.0 V, -0.1 V, and -0.2V, respectively.

**Fig. 4:** Spin conductance polarization $\alpha = (G_{up} - G_{down})/(G_{up} + G_{down})$ versus $V_{sweep}$ of a dual QPC with the following parameters: $w_1 = 48$ nm, $l_1 = 80$ nm, $w_2 = 16$ nm, $l_2 = 22$ nm, $d = 4$ nm. The different curves correspond to the biasing conditions $V_{sg1} = -0.2$ V $+ V_{sweep}$, $V_{sg2} = +0.2 + V_{sweep}$ and $V_{sg3} = V_{sg4} = 0.0$ V, $-0.1$ V, and $-0.2$V, respectively. Also shown for comparison is $\alpha$ vs. $V_{sweep}$ for a QPC with a single pair of in-plane gates and a channel length of 48nm.

$V_{sg1}$
$+V_{sweep}$
$V_{sg3}$
d

$V_s$
I
II
$W_2$
$L_2$
$L_2$
$V_d$
$W_1$

$V_{sg2}$
$+V_{sweep}$
$V_{sg4}$

$L_1$

**Figure 1 (Charles et al.)**

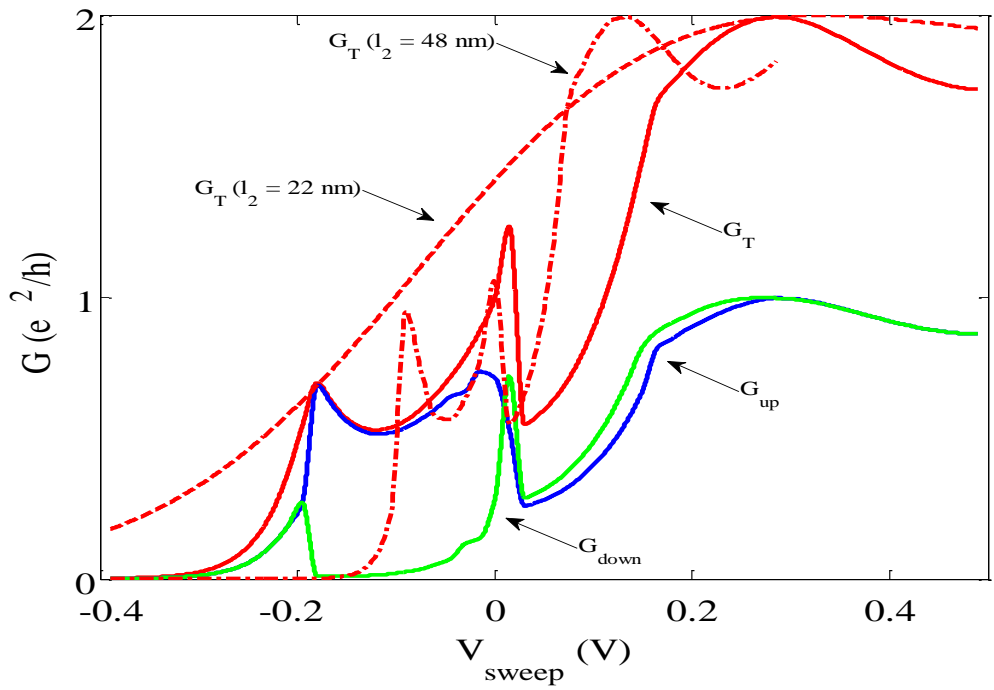

**Figure 2 (Charles et al.)**

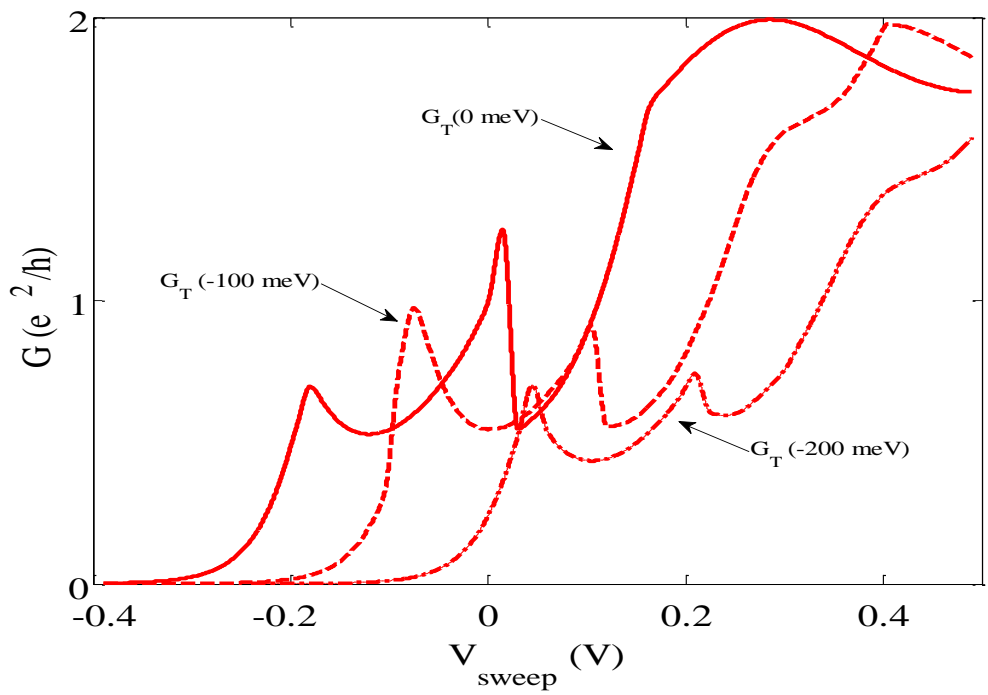

**Figure 3 (Charles et al.)**

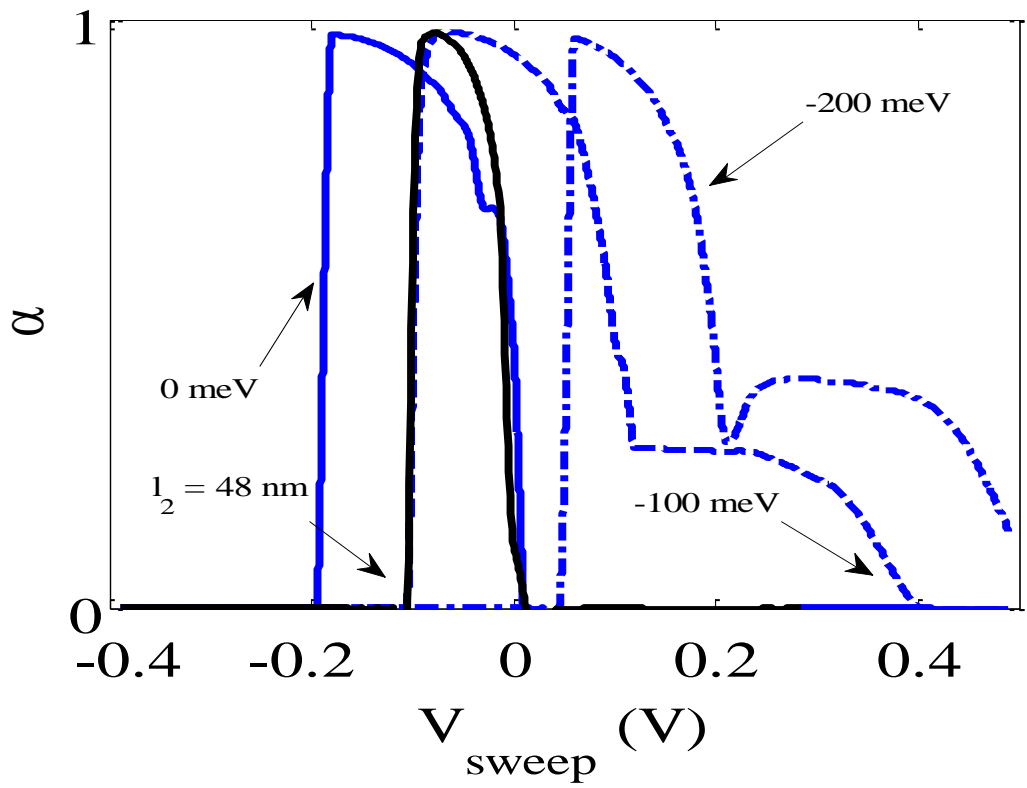

**Figure 4 (Charles et al.)**